\documentclass[10pt,aps,prd,twocolumn,floats,floatfix,showpacs,superscriptaddress,nofootinbib]{revtex4-1}
\usepackage[utf8]{inputenc}               		

\usepackage{graphicx,mathtools,amssymb,amsmath,amsthm,amsfonts,epsfig,epsf}
\usepackage[outdir=./]{epstopdf}
\usepackage[linktocpage]{hyperref}
\hypersetup{pdfstartview=}
\usepackage[usenames]{color}		
\usepackage{tensor}	
\usepackage{csquotes}
\usepackage{mathrsfs}
\usepackage[caption=false]{subfig}

\newcommand{\be}{\begin{equation}}
\newcommand{\ee}{\end{equation}}
\newcommand{\pt}{\partial}
\newcommand{\beq}{\begin{eqnarray}}
\newcommand{\eeq}{\end{eqnarray}}
\begin{document}

\title{Primordial black holes from an inflationary valley}

\author{Bao-Min Gu}
\email{gubm@ncu.edu.cn}
\affiliation{Department of physics, Nanchang University,
Nanchang 330031, China}
\affiliation{Center for Relativistic Astrophysics and High Energy Physics, Nanchang University, Nanchang,
330031, China}

\author{Fu-Wen Shu}
\email{shufuwen@ncu.edu.cn}
\affiliation{Department of physics, Nanchang University,
Nanchang 330031, China}
\affiliation{Center for Relativistic Astrophysics and High Energy Physics, Nanchang University, Nanchang,
330031, China}
\affiliation{Center for Gravitation and Cosmology, Yangzhou University, Yangzhou, China}

\author{Ke Yang}
\email{keyang@swu.edu.cn}
\affiliation{School of Physical Science and Technology, Southwest University, Chongqing 400715, China}

\author{Yu-Peng Zhang}
\email{zyp@lzu.edu.cn}
\affiliation{Lanzhou Center for Theoretical Physics, Key Laboratory of Theoretical Physics of Gansu Province, School of Physical Science and Technology, Lanzhou University, Lanzhou 730000, China}
\affiliation{Institute of Theoretical Physics, Research Center of Gravitation, Lanzhou University, Lanzhou 730000, China}

%%%

\begin{abstract}
Primordial black holes (PBHs) could be formed if large perturbations are generated on small scales in inflation. We study a toy inflation model with a local minimum. The curvature perturbations are enhanced when the inflaton passes through the local minimum, with more efficient amplification rate than that of quasi-inflection point inflation, leading to the production of PBHs on small scales. The PBHs could comprise the total dark matter in the mass window $10^{16}$--$10^{20}$g.
\end{abstract}

\maketitle

%%%
\section{Introduction}

Primordial black holes (PBHs) are hypothetical objects formed in early Universe, proposed by Zel'dovich and Novikov \cite{Zeldovich1966}, and developed by Hawking and Carr \cite{Hawking:1971ei,Carr:1974nx,Carr:1975qj}. It is believed that they could be formed in regions with large density perturbations. The most important motivation to study PBHs is that they are natural dark matter candidate. Although latest observations imposed tight constraints on the abundance of PBHs \cite{Khlopov:2008qy,Carr:2009jm,Carr:2016drx,Carr:2017jsz,Poulin:2017bwe,Sasaki:2018dmp,Carr:2020gox,Green:2020jor}, the PBHs that have not evaporated could contribute a fraction in wide mass ranges. In particular, there are still windows, e.g. $10^{16}\text{g}<M_\mathrm{PBH}<10^{20}\text{g}$, in which the PBHs could constitute a significant fraction of dark matter. In addition to dark matter, PBHs are also the possible origin of black holes that cannot be formed in standard gravitational collapse processes, for example, the intermediate mass black holes. According to the supernova explosion theory \cite{Barkat:1967zz,Heger:2001cd,Woosley:2007qp,Woosley:2016hmi,Farmer:2019jed},
black holes with mass around $50M_{\odot}-140 M_{\odot}$ cannot be formed due to the pair instability.
However, in GW190521 \cite{LIGOScientific:2020iuh,LIGOScientific:2020ufj} at least one of the black holes falls in the predicted mass gap,
bringing a serious challenge to the theory of astrophysical black hole formation. PBHs provide a possible explanation for the black holes in GW190521, as long as the PBHs accrete efficiently \cite{DeLuca:2020sae,Chen:2021nxo}.

To produce PBHs with observationally interesting abundance in inflation, the primordial curvature perturbations and the power spectrum should be large enough. Roughly speaking, one needs a power spectrum as large as $\mathcal{O}(10^{-2})$ on scales that are much smaller than that of cosmic microwave background (CMB). Therefore, even the nondetection of PBHs imposes constraints on the early Universe. In slow roll inflationary models, PBHs cannot be formed because the power spectrum is nearly scale invariant and keeps small on all scales. It was considered in \cite{Garcia-Bellido:2017mdw,Ezquiaga:2017fvi,Germani:2017bcs,Motohashi:2017kbs} that inflation models with a quasi-inflection point, known as the ``ultra slow roll" (USR) inflation, could amplify the curvature perturbations and  generate PBHs on small scales. This kind of potential includes a rather flat region in which the slope $V'(\phi)\simeq 0$. When inflaton passes though the plateau, the slow roll parameter $\epsilon$ decreases exponentially and $\eta\simeq-6$. This makes the curvature perturbations and the power spectrum grow exponentially. The success of this kind of model relies on fine-tuning of the theory parameters, so that the potential is flat enough to produce PBHs on small scales and be consistent with the CMB observations on large scales.
Nevertheless, one should note that the ``deceleration" of inflaton is not necessary to amplify the curvature perturbations. Instead, ``acceleration" of the inflaton may also amplify the perturbations. In Refs. \cite{Kefala:2020xsx,Dalianis:2021iig,Inomata:2021uqj,Inomata:2021tpx,Cai:2021zsp,Boutivas:2022qtl}, the authors provide some examples of inflation with step features that seem to accelerate the inflaton and suppress the power spectrum. However, the perturbations can still be amplified due to the USR phase right after the acceleration, provided that the width and height of the steps are appropriate. Reference \cite{Inomata:2021tpx} also pointed out that steps of smooth functions would lead to strong coupling of the perturbations during transition, thus only piecewise functions could  generate PBHs successfully. Another  mechanism of interesting is the sound speed resonance \cite{Cai:2018tuh,Chen:2020uhe}, in which the sound speed oscillates so that the perturbations possess resonance effect and the power spectrum is enhanced on certain scales. Besides, one may also consider inflation with bumps \cite{Mishra:2019pzq,Zheng:2021vda},  nonstandard theories like modified gravity \cite{Barrow:1996jk,Drees:2019xpp,Fu:2019ttf,Kawai:2021edk}, inflation driven by noncanonical field \cite{Ozsoy:2018flq,Kamenshchik:2018sig,Lin:2020goi,Yi:2020cut,Gao:2020tsa,Solbi:2021wbo,Gao:2021vxb},  inflation with intermediate noninflationary phase \cite{Fu:2020lob}, etc.

In this work, however, we study a nonmonotonous potential with a local minimum. The inflaton starts in slow roll phase and then passes through the local minimum. The inflaton could ``climb out" the potential valley classically as long as the valley is not too deep. The amplification mechanism of this work is similar to that of USR inflation, but with a more efficient amplification rate. This idea has been considered in previous literature \cite{Ozsoy:2018flq,Kannike:2017bxn,Ballesteros:2017fsr,Hertzberg:2017dkh,Biagetti:2018pjj,Figueroa:2020jkf}. In this work we give a concrete realization without the use of piecewise functions and approximation.

The paper is organized as follows. In section \ref{section 1} we review the conditions of enhancing the power spectrum. In section \ref{model setup}, we present our model and the setup. In section \ref{PBH formation} we give the main results of this paper and show that our model could amplify the curvature perturbations sufficiently and produce PBHs. Finally, we conclude in section \ref{conclusion} with a short discussion on this work.

\section{The conditions of enhancing the curvature perturbations}
\label{section 1}
We start from the linearized FLRW background spacetime,
\beq
\mathrm{d}s^2=&&-(1+2\Psi)\mathrm{d}t^2
+2a(t)\pt_i\xi\mathrm{d}t\mathrm{d}x^i \nonumber \\
&&+a^2(t)\left[\delta_{ij}(1-2\Phi)+\pt_i\pt_j E\right]\mathrm{d}x^i\mathrm{d}x^j.
\label{pert metric}
\eeq
The inflaton field is decomposed into a homogeneous background field and a small inhomogeneous fluctuation, i.e., $\phi(t,\mathbf{x})=\phi(t)+\delta\phi(t,\mathbf{x})$.
The background evolution of the inflaton is governed by
\be
\ddot{\phi}+3H\dot{\phi}+V_{,\phi}=0.
\ee
It is usually more convenient to work with the e-folding number $N$ in numerical calculations. Using the transformation relation between $t$ and $N$
\be
H\mathrm{d}t=\mathrm{d}N,
\ee
the evolution equation becomes
\be
\phi_{,NN}+3\phi_{,N}-\frac{1}{2}\phi_{,N}^{3}
+\left(3-\frac{1}{2}\phi_{,N}^{2}\right)\frac{V_{,\phi}}{V}=0,
\label{beq evolution}
\ee
where $\phi_{,N}=\frac{\partial \phi}{\partial N}$ and $\phi_{,NN}=\frac{\partial^2 \phi}{\partial N^2}$, and the Planck mass is set to be 1 in this equation.

The primordial fluctuations can be characterized by the gauge invariant curvature perturbation,
\be
\mathcal{R}=\Phi+\frac{H}{\dot{\phi}}\delta\phi.
\ee
In terms of the e-folding number $N$, the evolution equation of the curvature perturbation in momentum space can be written as
\be
\mathcal{R}_{k,NN}+(3-\epsilon+\eta)\mathcal{R}_{k,N}
+\frac{k^2}{a^2 H^2}\mathcal{R}_k =0,
\label{curvature evolution}
\ee
where the slow roll parameters are defined by
\beq
\epsilon&\equiv&-\frac{\dot{H}}{H^2}=\frac{1}{2}\phi_{,N}^2,\\
\eta&\equiv&\frac{\dot{\epsilon}}{H\epsilon}=2\frac{\phi_{,NN}}{\phi_{,N}}.
\eeq
Therefore, the evolution of the curvature perturbation $\mathcal{R}_{k}$ can  be principally determined by Eqs. (\ref{beq evolution}) and (\ref{curvature evolution}).
In the slow roll phase, the slow roll parameters are tiny and
the curvature perturbation $\mathcal{R}_k$ is frozen on super-Hubble scales, i.e., $k\ll aH$.
Using the solution of $\mathcal{R}_k$, the scalar power spectrum can be expressed as
\be
\mathcal{P}_\mathrm{s}(k)=\frac{k^3}{2\pi^2}|\mathcal{R}_{k}|^2.
\label{powerspectrum}
\ee
For slow roll inflation, the power spectrum can be well approximated as
\be
\mathcal{P}_\mathrm{s}(k)=\frac{H^2}{8\pi^2\epsilon}\bigg|_{k=aH}.
\label{SRpowerspectrum}
\ee
It is calculated at the time when the CMB pivot scale exits  the horizon ($k=aH$). But this approximation breaks down if the slow roll conditions are violated.

For more general cases this approximation may not be accurate. In USR inflation, the potential is adjusted to be extremely flat, i.e., $V_{,\phi}\simeq 0$ in a short range, so that the potential has a quasi-inflection point.   When the inflaton rolls through this region, the slow roll parameter $\epsilon$ decreases exponentially and $\eta\simeq -6$. Hence the coefficient $3-\epsilon+\eta\simeq-3$. In USR phase, curvature perturbations would oscillate with increasing amplitude before horizon exit and evolve approximately as
\be
\mathcal{R}_k\propto e^{(-3+\epsilon-\eta)N}\simeq e^{3N}
\label{growingCurvature}
\ee
on super horizon scales. The amplification lasts until the slow roll conditions are recovered and the perturbations become frozen. In this case the power spectrum should not be evaluated at horizon exit. Instead, it should be calculated at the time after which the curvature perturbation is frozen. Thus the approximation (\ref{SRpowerspectrum}) would give incorrect predictions.
The amplification of curvature perturbation leads to an enhancement of the power spectrum. Generally, as long as the coefficient $3-\epsilon+\eta<0$, the curvature perturbation $\mathcal{R}_k$ would have exponentially growing mode.
This is certainly not enough to produce PBHs, since the formation of PBHs requires an enhancement of power spectrum at least seven orders of magnitude. Therefore, the amplification of curvature perturbation should be rapid and long enough.  Usually, the USR inflation models require fine-tuning of parameters.

In the following, we propose a toy inflation model with a small valley. Generally, inflation would end if the potential has a local minimum. However, we show that this is not the case if the potential is carefully constructed.

\section{The model}
\label{model setup}
The model we considered in this paper is constructed by a combination of potentials,
\begin{figure}[htb]
  \centering
  \includegraphics[width=6.5cm]{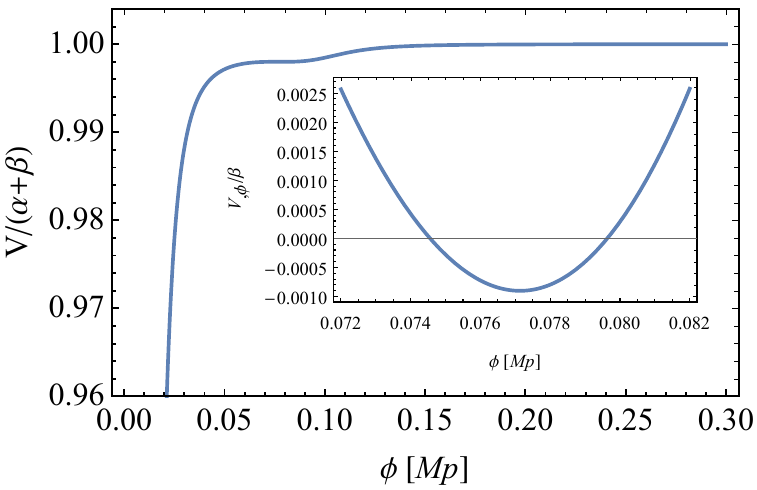}
  \includegraphics[width=6.50cm]{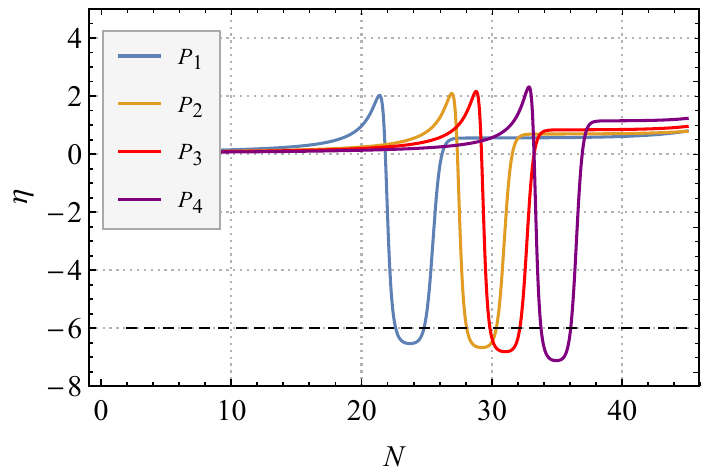}
  \caption{The configuration of the potential $V(\phi)$ (upper panel) with parameters given by the $P_4$ set in Table. \ref{table1}), and the numerically calculated slow roll parameter $\eta$ (lower panel). The inset in  the upper panel shows the slope of the potential around the local minimum.}
  %This clearly shows that $\eta<-6$ around the bottom of the potential valley.
  \label{figure1}
\end{figure}
\be
V(\phi)=\alpha \frac{V_1(\phi)}{1+V_1(\phi)}+\xi V_2(\phi),
\label{inflationpotential}
\ee
where $V_1=\frac{1}{4}\lambda \phi^2\left((\phi-\upsilon_1)^2+\upsilon_2^2\right)$ with $\lambda=4.0\times 10^6~M_p^{-4}$. The first part $\alpha V_1(\phi)/(1+V_1(\phi))$ is tuned such that it has a local minimum apart from the global minimum, see Fig. \ref{figure1}.
The second part $\xi V_2(\phi)$ provides slow roll inflation and behaves as the base potential.  The depth of the local minimum should be small enough so that the inflaton has enough kinetic energy to ``climb out." When the inflaton rolls down to the local minimum from slow roll phase, the slow roll parameter $\epsilon$ increases by several magnitudes and then decreases exponentially like that of USR phase. After it passes through the minimum, the slope $V_{,\phi}$ changes its sign and $\epsilon$ continues decreasing exponentially, with $\eta<-6$. In such a way, the coefficient $3-\epsilon+\eta<-3$, which makes the curvature perturbations grow faster than $e^{3N}$, and the power spectrum grows as
\be
\mathcal{P}_s\sim e^{2(-3-\eta)\Delta N},
\ee
where $\Delta N$ is the duration of the extremely slow roll phase, and we neglected the slow roll parameter $\epsilon$ since it is small. For USR inflation, $\eta\simeq-6$ and $\mathcal{P}_s\sim e^{6\Delta N}$. One needs at least $\Delta N\simeq 2.7$ to get seven orders of amplification. In our model, $\eta\simeq -7$ and $\mathcal{P}_s\sim e^{8\Delta N}$ for parameter set $P_4$ in Tab. \ref{table1}, thus one needs only $\Delta N\sim 2.1$ to produce PBHs. We see that our model provides a more efficient way to enhance the power spectrum and may reduce the fine tuning of parameters. For a more detailed analysis of the power spectrum, see Refs. \cite{Byrnes:2018txb,Cheng:2018qof,Ng:2021hll,Karam:2022nym}.
This extremely slow roll phase lasts until the inflaton ``climbs out" the potential valley.

\begin{figure}[htb]
  \centering
  \includegraphics[width=7cm]{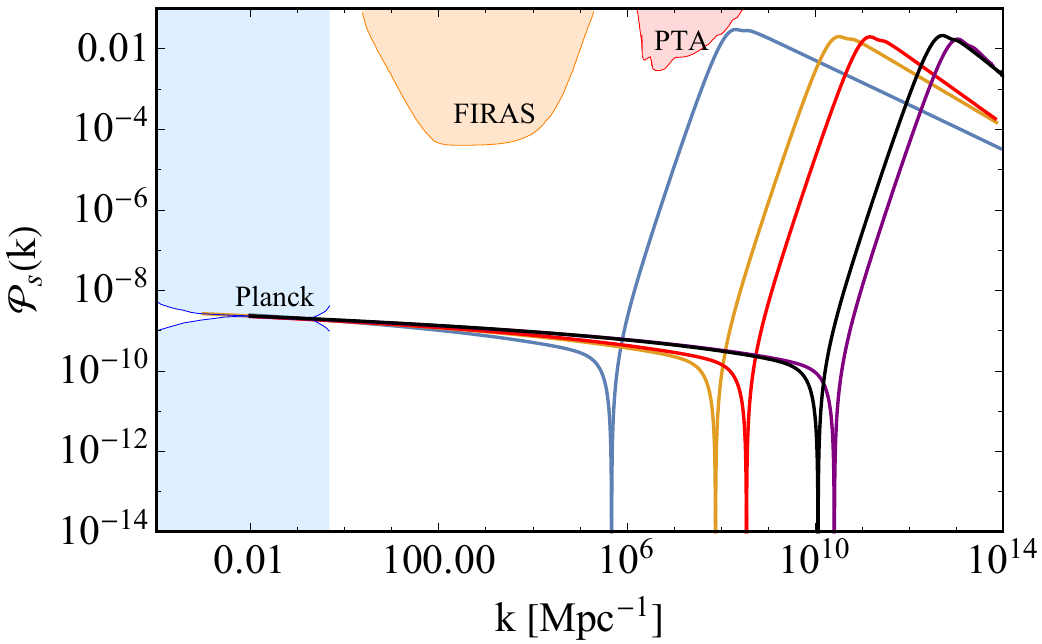}
%  \hfill
%  \includegraphics[width=7.0cm]{potentialV.pdf}
  \caption{Numerically calculated power spectrum by Bingo \cite{Hazra:2012yn}, for parameter sets $P_1$ (blue), $P_2$ (orange), $P_3$ (red), $P_4$ (purple), and $P_5$ (black) given in Tab. \ref{table1}.  The bounds are from Planck \cite{Planck:2018jri}, COBE FIRAS data (FIRAS) \cite{Fixsen:1996nj}, and Pulsar Timting Arrays (PTA) \cite{Byrnes:2018txb}.}
  \label{figure2}
\end{figure}
As a concrete example, we consider
\be
V_2(\phi)=\frac{(\phi/\upsilon_3)^4}{1+(\phi/\upsilon_3)^4}.
\ee
We have another 5 parameters, $\alpha$, $\beta$, $\upsilon_1$, $\upsilon_2$, $\upsilon_3$. First, we fix the ratio $\alpha/\beta$ to be 0.0125. Note that this ratio is not unique and other choices are also allowed. Next,
to be consistent with CMB observations \cite{Planck:2018jri}, we parameterize the potential with
\be
\mathcal{P}_s(0.05~\mathrm{Mpc}^{-1})\simeq \frac{V(\phi_{\mathrm{CMB}})}{24\pi^2 \epsilon(\phi_{\mathrm{CMB}})} = 2.1\times 10^{-9},
\ee
where $\phi_{\mathrm{CMB}}$ is the value of $\phi$ at which the  CMB pivot scale ($k=0.05~\mathrm{Mpc}^{-1}$) crosses the horizon.
The value of $\phi_{\mathrm{CMB}}$ is determined by requiring at least 50 e-folding numbers before the end of inflation.
Using the fact that $V(\phi_\mathrm{CMB})\simeq \alpha+\xi$, the parameters $\alpha$ and $\xi$ are constrained.  Furthermore, there is a constraint on the remaining parameters $\upsilon_i$ ($i=1,2,3$) from the spectral index,
\be
n_s-1=-6\epsilon_V+2\eta_V\simeq 2\eta_V,
\ee
where the approximately equal sign results from the hierarchy between the slow roll parameters. One may use the other slow roll parameters to give further restrictions on the model parameters, but there are always free parameters if we abandon the assumptions on ($\lambda$, $\alpha$, $\xi$), hence the parameter space cannot be severely constrained. We give some parameter sets and the corresponding spectral index and tensor-to-scalar ratio predicted in Tab. \ref{table1} as examples.
%\begin{center}
%\begin{table}[ht] % title of Table % used for centering table
%\begin{tabular}{c| c  c  c  c  c c c} % centered columns (6 columns)
%\hline\hline
\begin{table*}
\centering
\caption{Some model parameter sets and the predicted spectral index and tensor-to-scalar ratio.}
\begin{tabular*}{\textwidth}{@{\extracolsep{\stretch{1}}}*{9}{r}@{}}
\hline\hline
Sets &$\alpha [M_p^4]$&$\xi [M_p^4]$&  $\upsilon_1[M_p]$& $\upsilon_2[M_p]$ & $\upsilon_3[M_p]$
&$\phi_\mathrm{CMB}[M_p]$ & $n_s$ & $ r $
 \\ [0.5ex] %
\hline
$P_1$ & $3.98793\times 10^{-15}$ & $3.19035\times 10^{-13}$ &  0.098 & 0.02753647 & 0.0098& 0.193 & 0.9591 & $1.0\times 10^{-5}$ \\

$P_2$ & $2.66951\times 10^{-15}$ & $2.13561\times 10^{-13}$ &  0.097 & 0.02733370 & 0.0097& 0.198 & 0.9680 & $6.9\times 10^{-6}$\\

$P_3$ & $2.37415\times 10^{-15}$ & $1.89932\times 10^{-13}$ &  0.096 & 0.02711904 & 0.0096& 0.199 & 0.9700 & $6.2\times 10^{-6}$\\

$P_4$ & $1.88514\times 10^{-15}$ & $1.50833\times 10^{-13}$ &  0.094 & 0.02665191 & 0.0094& 0.201 & 0.9739 & $4.9\times 10^{-6}$\\

$P_5$ & $1.44931\times 10^{-15}$ & $1.44931\times 10^{-13}$ &  0.094 & 0.02446018 & 0.0099& 0.192 & 0.9731 & $4.7\times 10^{-6}$\\
 [0.5ex] % [1ex] adds vertical space %inserts single line
\hline
\end{tabular*}
\label{table1}
\end{table*}
%\end{tabular}
%\label{table1}
% % is used to refer this table in the text
%\end{table}
%\end{center}

\section{PBH formation}
\label{PBH formation}

The formation of PBHs requires the power spectrum to be at least $\mathcal{O}(0.01)$. This condition can be easily satisfied in our model, as can be seen from Fig. \ref{figure2}. Note that the power spectrum has broad peak and without oscillations. The curvature perturbation is related to the density contrast by
\be
\delta(t,k)=\frac{2(1+\omega)}{5+3\omega}\left(\frac{k}{aH}\right)\mathcal{R}_{k}.
\ee
Due to the amplification of the curvature perturbations, the overdense region ($\delta>\delta_c$) of the Universe would collapse to be PBHs when the perturbations reenter the horizon during radiation dominated era, with $\omega=\frac{1}{3}$ and $\delta(t,k)=\frac{4}{9}\mathcal{R}_{k}$.  The exact value of the threshold $\delta_c$ is uncertain since the details of the PBH formation process are not clear \cite{Kehagias:2019eil,Musco:2020jjb}.
Assuming a Gaussian probability distribution function of perturbations, the PBH mass fraction $\beta$ at formation time can be calculated as
\beq
\beta(M_{\mathrm{PBH}})&=&2\gamma\int_{\delta_c}^{1} \frac{\mathrm{d}\delta}{\sqrt{2\pi}\sigma_{M_\mathrm{PBH}}}\mathrm{exp}
\left(-\frac{\delta^2}{2\sigma^2_{M_\mathrm{PBH}}}\right)\nonumber\\
&\simeq&\sqrt{\frac{2}{\pi}}\frac{\gamma}{\nu_c}\mathrm{exp}\left(-\frac{\nu_c^2}{2}\right),
\eeq
where the constant $\gamma$ is fraction of mass transformed to be PBHs that has $\delta>\delta_c$ and we use $\gamma=0.2$ \cite{Carr:1975qj} in this paper.
$\nu_c=\delta_c/{\sigma_{M_\mathrm{PBH}}}$ with the variance
$\sigma_{M_\mathrm{PBH}}$  defined by
\be
\sigma_{M_\mathrm{PBH}}^2=\int_{0}^{\infty}\frac{\mathrm{d}k}{k}\frac{16}{81}(kR)^4W^2(kR)\mathcal{P}_s(k),
\ee
where $W(kR)$ is the window function used to smooth the density contrast on comoving scale $R$. We use the Gaussian type window function in this work,
\beq
W_\text{g}(kR)&=&\mathrm{exp}\left(-\frac{k^2R^2}{2}\right).
\eeq
One can also choose the top-hat type \cite{Lyth:2009zz}, which would lead to slightly different mass fraction \cite{Young:2019osy}. Finally, the mass fraction is obtained \cite{Young:2014ana},
\be
\beta(M_\text{PBH})\simeq\gamma\sqrt{\frac{2}{\pi}}\frac{4\sqrt{\mathcal{P}_\mathrm{s}(k)}}
{9\delta_c}\mathrm{exp}
\left(-\frac{81\delta_c^2}{32\mathcal{P}_\mathrm{s}(k)}\right).
\ee

Regarding the PBHs as a fraction of dark matter, the mass fraction can be related to the abundance $f_\text{PBH}$ by
\begin{figure}[htb]
  \centering
  \includegraphics[width=7.5cm]{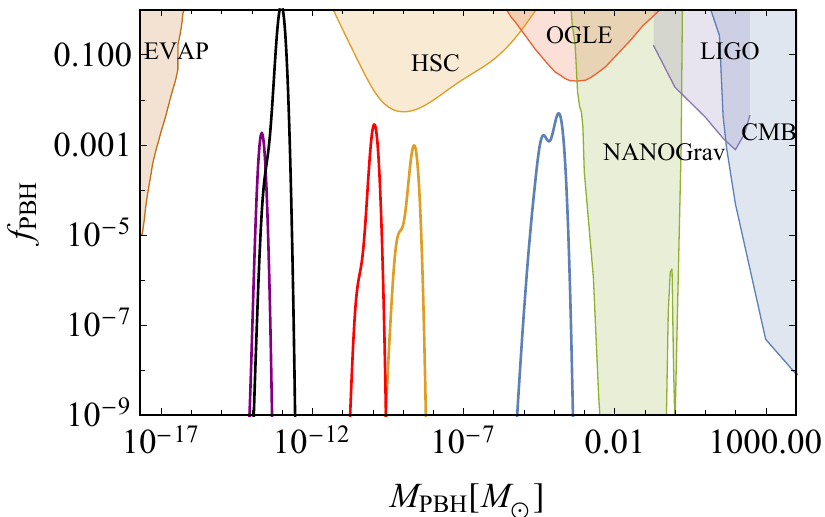}
%  \hfill
%  \includegraphics[width=7.0cm]{potentialV.pdf}
  \caption{The fraction of PBHs as dark matter for parameter sets $P_1$ (blue), $P_2$ (orange), $P_3$ (red), $P_4$ (purple), and $P_5$ (black) in Tab. \ref{table1}. The upper bounds are from evaporation of PBHs due to Hawking radiation (EVAP) \cite{Carr:2009jm,Clark:2016nst,Clark:2018ghm,Boudaud:2018hqb,Laha:2019ssq,DeRocco:2019fjq,Laha:2020ivk}, microlensing of stars in M31 by Subaru Hyper Suprime-Cam (HSC) \cite{Niikura:2017zjd}, Optical Gravitational Lensing Experiment (OGLE) \cite{Niikura:2019kqi}, North American Nanohertz Observatory for Gravitational waves (NANOGrav) \cite{Chen:2019xse}, LIGO/Virgo's search for subsolar mass ultracompact objects \cite{Kavanagh:2018ggo,LIGOScientific:2019kan}, and CMB \cite{Serpico:2020ehh}.}
  \label{figurebound}
\end{figure}
\beq
\beta(M_{\text{PBH}})&\simeq&3.7\times 10^{-9}\left(\frac{\gamma}{0.2}\right)^{-1/2}\nonumber\\ &\times&
\left(\frac{g_{*,\text{form}}}{10.75}\right)^{1/4}
\left(\frac{M_{\text{PBH}}}{M_{\odot}}\right)^{1/2}f_\text{PBH},
\eeq
where $g_{*,\text{form}}$ is the relativistic degrees of freedom at formation, and $M_{\odot}$ is the solar mass. Using the power spectrum and the relation between the mass of the formed PBH and the wave number,
\beq
M_\text{PBH}\simeq 30M_{\odot}\left(\frac{\gamma}{0.2}\right)\left(\frac{g_{*,\text{form}}}{10.75}\right)^{-1/6}
\nonumber\\
\times\left(\frac{k}{2.9\times 10^5 \text{Mpc}^{-1}}\right)^{-2},
\eeq
we get the abundance of PBHs as dark matter $f_{\text{PBH}}$.
For $\delta_c=0.51$ \cite{Young:2019osy}, the results are shown in Fig. \ref{figurebound}.
The predicted abundances of PBHs are safely below the constraints from observations for the parameter sets given in Tab. \ref{table1}.

We see that the mass of produced PBHs increases for larger values of $\upsilon_1$. This is because the position of the local minimum depends linearly on $\upsilon_1$, and longer wave length modes are amplified for large $\upsilon_1$, thus the power spectrum peaks at smaller $k$, producing heavier PBHs.
According to Tab. \ref{table1}, the parameter set $P_4$ produces the lightest PBHs, with the spectral index $n_s$ on the edge of the allowed values from CMB observations \cite{Planck:2018jri}. Indeed, for parameters producing lighter PBHs, the predicted spectral index $n_s$ has increasing tension with observations. This is mainly because of our parameter setup, $\alpha=0.0125\xi$ and $\lambda=4.0\times 10^{6} M_p^{-4}$. But the tension could be reduced if we relax the constraints on $\alpha$, $\xi$, and $\lambda$, since there are always free parameters. We give a parameter set $P_5$ with $\lambda=5.0\times 10^6 M_p^{-4}$ and $\alpha=0.01\xi$. This set produces the PBHs that could contribute 100\% of the dark matter.

At last, it should be noted that for some parameter spaces the peak of the power spectrum would be $\mathcal{O}(0.1)$ due to the rapid amplification. One needs to avoid these cases since it would make the linear theory break down and the above predictions untrustable. In fact, it is noticed that the nonlinear effects of perturbations may be important for the PBH abundance because of the exponential dependence of the mass fraction on the power spectrum \cite{Kawasaki:2019mbl,DeLuca:2019qsy,Young:2019yug}. In this work the nonlinear effects are not included since this issue is outside the scope of this paper.
\section{Conclusion}
\label{conclusion}
To summarize, we studied the possibility of amplifying the curvature perturbations on small scales and at the same time satisfying the CMB observations on large scales, by introducing a potential with local minimum. The amplification mechanism is similar to that of USR inflation but with a period  $\eta<-6$. We show that this type of inflation could produce PBHs successfully. As a toy model, it has some shortcomings. For example, the potential (\ref{inflationpotential}) introduces six parameters. To some extent this makes our theory seem not elegant since the large parameter space can not be severely constrained. In addition, the formalism of the potential is lack of explanation from fundamental physics. In spite of these shortcomings, this work mainly confirmed the idea of inflation with local minimum.  We expect to look for more elegant theory with less parameters, and with more natural formalism that can be derived from fundamental theory, so that the theory could be constrained by observations, for example the second order gravitational waves induced by the large perturbations \cite{Baumann:2007zm,Inomata:2016rbd,Kohri:2018awv,Cai:2018dig,Inomata:2018epa,Wang:2019kaf,Yuan:2021qgz}. We leave these considerations for future work.

\section{Acknowledgements}
This work is supported in part by the National Natural Science Foundation of China (Grants No. 12165013, No. 11975116, No. 12005174, and No. 12105126). K. Yang acknowledges the support of Natural Science Foundation of Chongqing, China under Grant No. cstc2020jcyj-msxmX0370. Y.-P Zhang acknowledges the Fundamental Research Funds for the Central Universities (Grant No. lzujbky-2021-pd08), the China Postdoctoral Science Foundation (Grant No. 2021M701531).

%\bibliographystyle{JHEP}
%\bibliography{D:/Work/JabFiles/allpaper}

\end{document}